\documentclass[twocolumn,showpacs,aps,prl,superscriptaddress]{revtex4}

\usepackage{graphicx}
\usepackage{dcolumn}
\usepackage{amsmath}
\usepackage{epsfig}
\usepackage{colordvi}
\usepackage{color}

\RequirePackage{xspace}

\usepackage{relsize}
\def\babar{\mbox{\slshape B\kern-0.1em{\smaller A}\kern-0.1em
    B\kern-0.1em{\smaller A\kern-0.2em R}}}

\mathchardef\Upsilon="7107
\def\Y#1S{\ensuremath{\Upsilon{(#1S)}}\xspace}
\def\FourS {\Y4S}
\def\pep2{PEP-II}

\newcommand{\BABARPubYear}    {03}
\newcommand{\BABARPubNumber}  {025}
\newcommand{\SLACPubNumber} {10203}

\def\figurebox#1#2#3{%
    \def\arg{#3}%
    \ifx\arg\empty
    {\hfill\vbox{\hsize#2\hrule\hbox to #2{\vrule\hfill\vbox to #1{\hsize#2\vfill}\vrule}\hrule}\hfill}%
    \else
    {\hfill\epsfbox{#3}\hfill}%
    \fi}

\begin{document}

\begin{flushleft}
\babar-PUB-\BABARPubYear/\BABARPubNumber\\
SLAC-PUB-\SLACPubNumber\\
\end{flushleft}

\title{
{\large \bf
{\boldmath $J/\psi$} production via initial state radiation in
{\boldmath $e^+e^-\to \mu^+\mu^-\gamma$}} 
at~an~{\boldmath $e^+e^-$}~center-of-mass energy near 10.6 GeV
}

%
\author{B.~Aubert}
\author{R.~Barate}
\author{D.~Boutigny}
\author{J.-M.~Gaillard}
\author{A.~Hicheur}
\author{Y.~Karyotakis}
\author{J.~P.~Lees}
\author{P.~Robbe}
\author{V.~Tisserand}
\author{A.~Zghiche}
\affiliation{Laboratoire de Physique des Particules, F-74941 Annecy-le-Vieux, France }
% ...
\collaboration{The \babar\ Collaboration}
\noaffiliation

\date{\today}

\begin{abstract}
We have used a study of the process  $e^+e^-\to \mu^+\mu^-\gamma $
at a center-of-mass energy near the \FourS\ resonance for a
$\mu^+\mu^-$ invariant mass range near the $J/\psi$ mass
to extract the cross section 
$ \sigma (e^+ e^- \to J/\psi\gamma \to \mu^+ \mu^-\gamma $).
The data set, corresponding to an integrated luminosity of 88.4 fb${}^{-1}$,
was collected using the \babar\ detector at the \pep2\ collider.
We measure the product
$\Gamma(J/\psi\to e^+ e^-)\cdot B(J/\psi\to\mu^+\mu^-) $
to be $ 0.330\pm0.008\pm0.007\mbox{ keV}$.
Using the world averages for $B(J/\psi\to\mu^+\mu^-)$ and
$B(J/\psi\to e^+ e^-)$,
we derive the $J/\psi$ electronic and  total widths:
$\Gamma(J/\psi\to e^+ e^-)=5.61\pm0.20\mbox{ keV}$ and
$\Gamma=94.7\pm4.4\mbox{ keV}$.
\end{abstract}

\vspace*{2mm}
\maketitle
The possibility of using $e^+e^-$ annihilation
with initial state radiation (ISR),
$e^+e^-\to \mbox{hadrons}+\gamma$, to measure
the $e^+e^-$ cross sections into hadrons over a wide range of
center-of-mass (CM) energies in a single experiment
has been discussed in the literature~\cite{ISR}.
In this paper, we have implemented this idea by studying the process
$e^+e^-\to\mu^+\mu^-\gamma$
for $\mu^+\mu^-$ masses in the range from 2.8 to 3.4 GeV/$c^2$.
We measure the cross section
$ \sigma (e^+e^- \to J/\psi\gamma \to \mu^+ \mu^-\gamma ) $
and derive the product
of electronic width times branching fraction
$\Gamma(J/\psi\to e^+ e^-)\cdot B(J/\psi\to\mu^+\mu^-)$.
Using the world averages for $B(J/\psi\to\mu^+\mu^-)$ and
$B(J/\psi\to e^+ e^-)$~\cite{pdg},
we then derive the electronic and  total widths of
the $J/\psi$ meson.
\begin{figure}[t]
\includegraphics[width=0.9\linewidth]{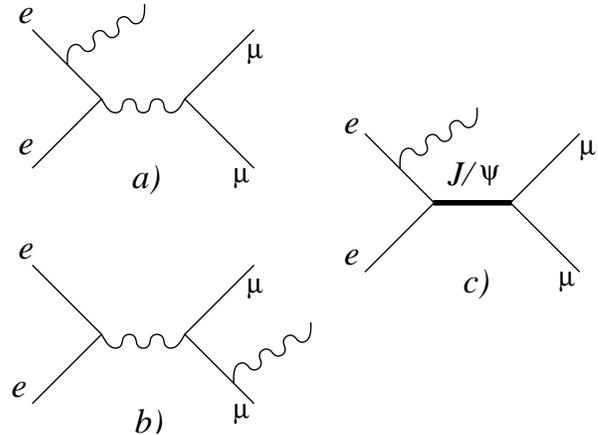}
\caption{Diagrams for $e^+e^-\to\mu^+\mu^-\gamma$.
a) Initial state radiation. b) Final state radiation.
c) $J/\psi$ production.}
\label{fei}
\end{figure}
The data used in this analysis were collected with 
the \babar\ detector~\cite{babar}
at the \pep2\ asymmetric $e^+e^-$ storage ring~\cite{pep}.

The Born cross section for the process $e^+e^-\to\mu^+\mu^-\gamma$ in
the $J/\psi$ mass region has contributions
from three Feynman diagrams, as illustrated in
Fig.~\ref{fei}.
The first and second of these diagrams
describe the pure QED processes
corresponding to initial state radiation and final state radiation (FSR).
The visible QED cross section in \babar~
 (defined by our ISR photon
acceptance) is about 1.2 pb in the di-muon mass range
2.8--3.4 GeV/$c^2$.
The contribution of the FSR process
to the QED cross section
depends on the photon energy and angle, and
is about 
10--20\% for the kinematic regime we study. 
The interference 
between ISR and FSR amplitudes does not change the total cross section,
but leads to charge asymmetries in the muon angular distributions.

The Born cross section for $J/\psi$ production (Fig.~\ref{fei}c) 
is given by
\begin{equation}
\frac{\mathrm{d}\sigma_{J/\psi}^{\mathrm{Born}} (s,x)}{\mathrm{d}x} = W(s,x)\cdot \sigma_0(s(1-x)),
\label{eq1}
\end{equation}
where
$\sqrt s $ is the $ e^+ e^- $ invariant mass,
$x  \equiv {2E_{\gamma}}/{\sqrt{s}}$, 
$E_{\gamma}$ is the photon energy in the CM,
and $\sigma_0$ is the Born cross section for 
$e^+e^-\to J/\psi\to\mu^+\mu^-$.
The function
\begin{equation}
 W(s,x) = \frac{2\alpha}{\pi x}\cdot (2\ln \frac{\sqrt{s}}{m_{e}}-1)
\cdot(1-x+\frac{x^2}{2})
\label{eq2} 
\end{equation}
describes the probability of ISR photon emission.
Here $\alpha$ is the fine structure constant and $m_{e}$ is the electron mass.
ISR photons are emitted predominantly at small angles relative to
the electron direction. About 10\% of the photons have CM
polar angles in the range $30^\circ<\theta<150^\circ$ and can be
detected in \babar.

As a first approximation,
the Born cross section for $e^+e^-\to J/\psi\to\mu^+\mu^-$ is given by the
Breit-Wigner formula
\begin{equation}
\sigma_0(s)=\frac{12\pi B_{ee} B_{\mu\mu}}{m^2}\cdot
\frac{m^2\Gamma^2}{(s-m^2)^2+m^2\Gamma^2},
\label{eq3}
\end{equation}
where $m$ and $\Gamma$ are the $J/\psi$ mass and total width, respectively,
$B_{ee}$ and $B_{\mu\mu}$ are the
branching fractions $B(J/\psi\to e^+ e^-)$ and $B(J/\psi\to \mu^+ \mu^-)$.
For a narrow resonance, such as the $J/\psi$,
we can replace the Breit-Wigner with a $\delta$ function 
$\pi m \Gamma\delta(s-m^2)$ and  integrate over photon energy
to find
\begin{equation}
\sigma_{J/\psi}^{\mathrm{Born}}(s)=\frac{12\pi^2 \Gamma_{ee} B_{\mu\mu}}{m\cdot s}\cdot
W(s,x_0);\: x_0=1-\frac{m^2}{s}.
\label{stot}
\end{equation}
Here $\Gamma_{ee}=\Gamma\cdot B_{ee}$.
\begin{figure}[t]
\includegraphics[width=\linewidth]{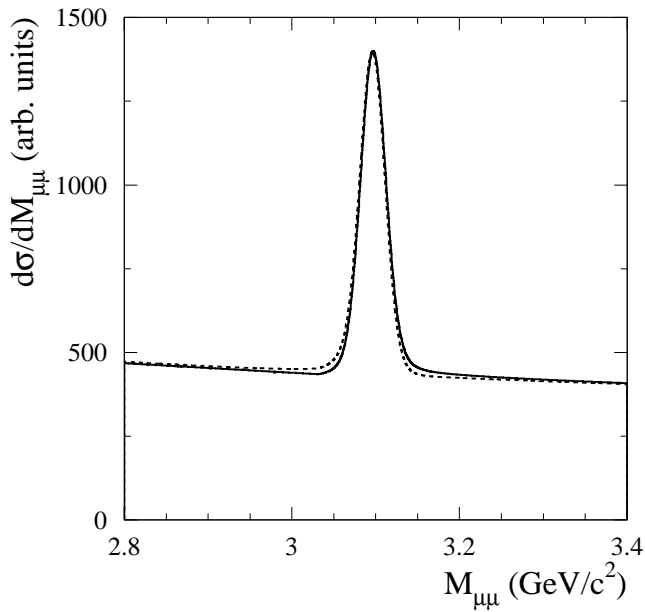}
\caption{The di-muon mass spectrum calculated with (solid line)
and without (dashed line) interference between the resonant $J/\psi$ 
production and QED amplitudes after convolution with the detector
resolution function.}
\label{inf1}
\end{figure}
These formulae do not account for interference
between the $e^+e^-\to J/\psi\to\mu^+\mu^-$ and the non-resonant (QED)
$e^+e^-\to \mu^+\mu^-$ amplitudes. 

The cross section
for $e^+e^-\to \mu^+\mu^-$ including QED
and resonant $ J/\psi $ production amplitudes
and their interference can 
be written  as~\cite{interf}
\begin{equation}
\sigma(s)=\frac{4\pi\alpha^2}{3s}
\biggl |1-Q \frac{m\Gamma}{m^2-s-i m\Gamma}\biggr |^2,
\label{interf}
\end{equation}
where $Q=3\sqrt{B_{ee}B_{\mu\mu}}/\alpha$.
The interference term changes sign at the $J/\psi$ mass.
Therefore, it does not change the integrated cross section 
of Eq.~(\ref{stot}) significantly,
but does change the shape of the mass distribution.
Because the $ J/\psi $ cross section is so much greater
than the QED cross section at resonance ($Q^2\approx 600$),
the power-law behavior of the Breit-Wigner tails produces
observable interference even 1000 widths from resonance.
The expected di-muon mass spectrum, convolved with
the detector resolution, is shown in Fig.~\ref{inf1}.
The interference is clearly seen, despite the experimental
resolution, 14.5 MeV/$c^2$, being
more than 100 times the
$J/\psi$ natural line width.
The maximum relative difference between the spectra calculated
with and without interference
is about 7\%.
The interference also leads to
a 1.3 MeV/$c^2$ shift between the maximum of the resonance peak and
the actual $J/\psi$ mass. 
The shape of the expected mass spectrum is
very sensitive to the tails of the Breit-Wigner approximation used in
Eqs.~(\ref{eq3}) and (\ref{interf}), where its
validity far from resonance is questionable.
To estimate the sensitivity of our analysis to the details of
the shape assumptions,
we will take the full difference between fits that do, and do not, use
interference as a measure of the systematic uncertainty.

The width of the $J/\psi$ has been measured directly in $p\bar{p}$ annihilation
with the result $99 \pm 12 \pm 6$ keV~\cite{pbarp}. In $e^+e^-$ annihilation,
measuring the area under the resonance curve for $e^+e^-\to J/\psi\to \mu^+ \mu^-$
gives the product $\Gamma_{ee} \cdot B_{\mu\mu}$
as seen in Eq.~(\ref{stot}). Combining this with the leptonic branching ratio
yields the total width. The BES Collaboration made a comprehensive
collection of measurements at the $J/\psi$ from which they determined
$\Gamma=84.4 \pm 8.9$ keV~\cite{BESwid}. This superseded
results obtained from original measurements made of the area under the
excitation curve in 1975. More recently, the BES Collaboration has
measured the leptonic branching ratio with a 1.5\% uncertainty using
$J/\psi$'s from the decay $\psi(2S)\to J/\psi \pi \pi$~\cite{BES1}.
It is this result that we combine with our measurement of
$\Gamma_{ee} \cdot B_{\mu\mu}$ to obtain the highest precision result
to date for the total width of the $J/\psi$.

Charged particle tracking for the \babar\ detector is provided by
a five-layer silicon vertex
tracker (SVT) and a 40-layer drift chamber (DCH), operating in
a 1.5\,T axial magnetic field. The transverse momentum
resolution is 0.47\% at 1 GeV/$c$. Energies of photons and
electrons are measured by a CsI(Tl) electromagnetic calorimeter (EMC)
with resolution of 3\% at 1 GeV.
Charged particle
identification is provided by ionization measurements
in the SVT and DCH, and by an internally reflecting ring-imaging
Cherenkov detector (DIRC).
Muons are identified in the
solenoid's instrumented flux return (IFR), which consists of
iron plates interleaved with
resistive plate
chambers.
The data sample used for this analysis corresponds to an
integrated luminosity of 88.4~fb$^{-1}$ recorded
 in the vicinity of the \FourS\ resonance.

The initial selection of $\mu^+\mu^-\gamma$ candidates
requires that all particles are detected inside a fiducial volume
and that the event kinematics are
consistent with the hypothesis $ e^+ e^- \to \mu^+\mu^-\gamma$.
Photons must have
polar angles in the range  $0.35 <\theta < 2.4$ radians and
must have a CM energy above 3 GeV.
Muon candidates must have polar
angles in the range
$0.35 <\theta < 2.4$ radians and
transverse momenta above 0.1 GeV/$c$,
and must originate from the interaction point.
Energy and momentum balance is provided
by the conditions $|E_{\rm total}-E_{\rm beams}|< 1.5 \mbox{ GeV}$
and $\Delta\Psi < 0.07 $.
Here $E_{\rm total}$ is the summed energy of the
muon candidates and the photon,
$\Delta\Psi$ is the angle between the photon and
the direction of the di-muon missing momentum,
$
 \bf{p}_{\rm miss} \equiv \bf{p}_{e^+} + \bf{p}_{e^-} - \bf{p}_{\mu^+}
- \bf{p}_{\mu^-}  \, .
$
We reduce backgrounds
using a one-constraint fit to the hypothesis 
that the recoil mass against the di-muon be zero.
Requiring $ \chi^2 < 20$
rejects 90\% of the multihadron ISR contamination (general $e^+e^-\to q 
\bar{q}\gamma$ reactions)
and about 10\% of signal events.

\begin{figure}[t]
\includegraphics[width=\linewidth]{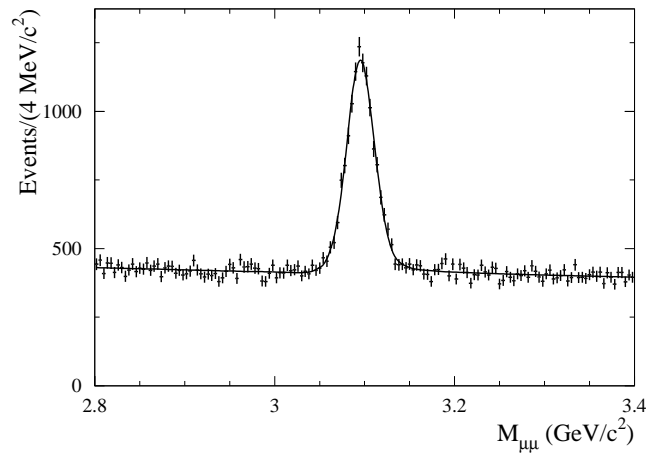}
\caption{
The mass spectrum for observed events. The curve is 
the result of the fit described in the text.}
\label{fit}
\end{figure}

The large background from
$e^+e^-\to e^+e^-\gamma$ is suppressed by
requiring that the charged track  momenta be greater than 0.5 GeV/$c$
and that the corresponding energies detected in the calorimeter be
small: $E_{\rm EMC} < 0.4\mbox{ GeV } $ for each track.
The average energy deposition of muons in the calorimeter is
about 0.2 GeV, while electrons typically deposit more than 90\%
of their energy in the calorimeter. 
Additional suppression of this
background is achieved by requiring 
large angular separation (in the CM) between the charged tracks and the photon
($\cos{\theta_{\mu\gamma}^*}<0.5$).
This also reduces the level of FSR $\mu^+\mu^-\gamma$
events in the final sample by a factor of two. 
The invariant mass distribution of approximately 70000
di-muon pairs in our final sample is shown in Fig.~\ref{fit}.
About 7800 events are in the $J/\psi$ peak.

To increase our detection efficiency and minimize systematic 
uncertainties, we do not use IFR information in selecting muons.
However, we do use this information for estimating backgrounds.
For this purpose, muon identification requires that
a track penetrate at least 2 nuclear interaction lengths ($\lambda$)
of IFR material, and that the difference between the measured and expected
muon ranges be less than $2 \lambda$.
This algorithm is 90\% efficient for true muons and
misidentifies about 10\% of real electrons as muons.

The remaining electron contamination in our final sample is
estimated using a subsample of events enriched with electrons.
We require that neither muon be identified in the IFR
using the algorithm described above.
We then require that the DCH based $dE/dx$  measurements for the
two tracks be consistent with the di-electron hypothesis and 
inconsistent with the di-muon hypothesis.
This eliminates 95\% of di-muons
and retains 85\% of di-electrons.
The fraction of electron events in the final 
sample is estimated to be $(0.1\pm0.1)\%$.

ISR events with hadronic final states are another
source of background, both on resonance and off.
For 
$e^+e^-\to J/\psi \gamma \to \pi^+ \pi^-\gamma $ and
$e^+e^-\to J/\psi \gamma \to K^+ K^-\gamma $,
the cross sections
are proportional to the ratios of branching fractions
$B(J/\psi\to \pi^+\pi^-)/B(J/\psi\to \mu^+\mu^-)\approx 2.5\cdot 10^{-3}$
and
$B(J/\psi\to K^+K^-)/B(J/\psi\to \mu^+\mu^-)\approx 4\cdot 10^{-3}$.
To first approximation, the off-resonance ratios,
$  \sigma (e^+ e^- \to \pi^+ \pi^-) /  \sigma (e^+ e^- \to \mu^+ \mu^-) $
and
$  \sigma (e^+ e^- \to K^+ K^-) /  \sigma (e^+ e^- \to \mu^+ \mu^-) $
are similar to those on resonance.
As off-resonance production proceeds via virtual photon intermediate
states while on-resonance production proceeds via both
virtual photon intermediate states and hadronic intermediate 
states~\cite{MilanaEtAl1993},
the on-resonance ratios are overestimates of the off-resonance ratios.
Thus, we consider the on-resonance background rate as
an upper limit for both.

The suppression of kaon and pion reactions
was studied using samples of
$e^+e^-\to\omega\gamma\to\ 3\pi\gamma$
and  $e^+e^-\to\phi\gamma\to K^+ K^-\gamma$
events.
About two thirds of these events are rejected
by the calorimeter energy deposition requirement.
Under the di-muon hypothesis,
the peak of the $J/\psi\to K^+K^-$ distribution
transforms into a broad distribution with
$M_{\mu\mu}<2.95$ GeV/$c^2$.
The only background peaking under the $ J/\psi 
\to \mu^+ \mu^- $ signal is that due to
$ J/\psi \to \pi^+ \pi^- $;
its contribution is
estimated to be $(0.09\pm0.03)\%$.
The only other decay into two charged hadrons, $J/\psi\to p\bar{p}$,
produces events with $M_{\mu\mu}<2.4$ GeV/$c^2$, and thus contributes 
no background in the
di-muon mass range studied here.
The total non-resonant background from
$e^+e^-\to e^+ e^-\gamma,\;\pi^+\pi^-\gamma,\;K^+ K^-\gamma$ processes 
is estimated to be $(0.3\pm0.2)\%$. 

The background from ISR production of higher multiplicity
multihadron events is estimated from Monte Carlo simulation.
We estimate the background from multihadron $J/\psi$ decays
to be less than 0.05\% using simulated
$e^+e^-\to J/\psi\, \gamma,\; J/\psi \to 3\pi$ events
and $J/\psi$ charged particle multiplicity data~\cite{BES1}.
We also note that such
events populate the mass region below 3 GeV/$c^2$
when misidentified as signal events.
We use the JETSET~\cite{Jetset} event generator to simulate the hadronic part of 
the $e^+e^-\to q \bar{q} \gamma,\: q=u,d,s$ cross section.
We find the background due to such events to be
less than 0.3\%. As these background rates are not the 
dominant sources of systematic uncertainty in our final results, 
we have not tried to determine them with greater precision.

\begin{figure}[t]
\includegraphics[width=\linewidth]{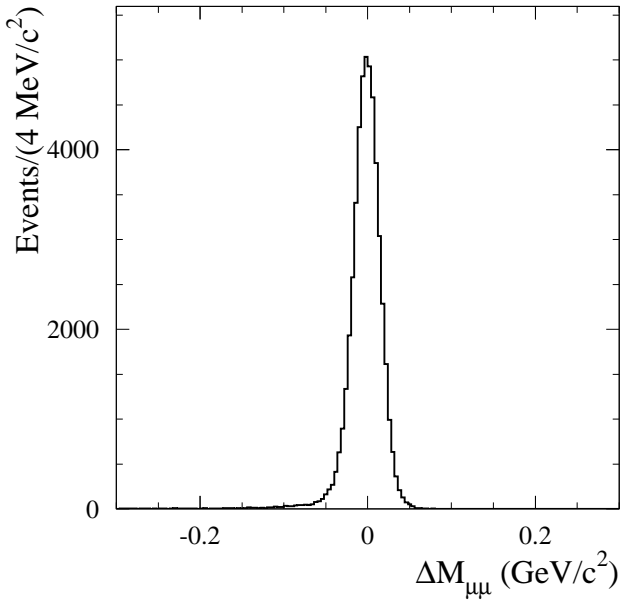}
\caption{
The distribution of
the reconstructed mass minus the generated mass in Monte Carlo events.}
\label{massres}
\end{figure}

We use a binned maximum likelihood fit to describe the mass
spectrum of Fig.~\ref{fit}.
The mass range used, 2.8--3.4 GeV/$c^2$, is divided into 150 bins of
width 4 MeV/$c^2$.
The probability density function (PDF) for the $J/\psi$ signal is
modeled as the convolution of a $J/\psi$ Breit-Wigner line shape and
the resolution function shown in Fig.~\ref{massres}.
This is
derived from detector simulation in conjunction with
an $e^+e^-\to\mu^+\mu^-\gamma$ event generator based
on the differential 
cross sections of Ref.~\cite{kuraev}.
Soft photon radiation is generated with the use of
the structure function method of Ref.~\cite{strfun}
and the PHOTOS package~\cite{PHOTOS} for electron and muon
bremsstrahlung, respectively. 
Muon bremsstrahlung
leads to the low mass tail observed in the
spectrum of Fig.~\ref{massres}. 
To account for possible
resolution differences between simulation and data,
the resolution function shown in Fig.~\ref{massres} is convolved with 
an additional Gaussian smearing
function of width $ \sigma_G$. 
Both $ \sigma_G $ and the observed
$J/\psi$ peak position, $M_{J/\psi}$,
are parameters in our fit. 
A Monte Carlo calculation shows
that the shape of the non-resonant cross section 
can be described well by a linear function. 
To account for
possible deviations from this hypothesis ({\em e.g.}, due to
detector response) a second-order polynomial is used to fit the experimental
spectrum. 
The full PDF is written as
\begin{eqnarray}
f(m_i)& = &\frac{N_0}{C(m_i)}[R\cdot H(m_i;M_{J/\psi},\sigma_G)+\nonumber \\
& &1+a (m_i-M_{J/\psi})+b (m_i-M_{J/\psi})^2],
\end{eqnarray}
where $m_i$ is the central value of the $i$th bin of the data histogram,
$N_0=\frac{\mathrm{d}N}{\mathrm{d}m}\cdot\Delta m$ is the level of
the non-resonant mass distribution at $m=M_{J/\psi}$,
$\Delta m=4\mbox{ MeV}/c^2$ is the bin width,
$H$ is the PDF for the $J/\psi$ signal with detector resolution,
and $a$ and $b$ are the background polynomial coefficients. 
To account for the interference between resonant and non-resonant
amplitudes described in Eq.~(\ref{interf}), the PDF is
divided by the correction function $C(m_i)$,
which is the ratio of the di-muon mass spectra
calculated with and without interference as shown in Fig.~\ref{inf1}.
Because the shape of this
function depends on the $J/\psi$ parameters (mass, full width),
an iterative procedure is used to calculate it. 
The ratio
\begin{equation}
R=\frac{N_{J/\psi}}{\frac{\mathrm{d}N}{\mathrm{d}m}\cdot \Delta m}
\end{equation}
is the main fit parameter. 
Here $N_{J/\psi}$ is the number of observed $J/\psi$ 
decays. 
After substituting cross sections for numbers of events,
this ratio can be rewritten
\begin{equation} 
R=\frac{\sigma_{J/\psi}^{\mathrm{Born}}}
{\frac{\mathrm{d}\sigma_{\mathrm{ISR}}^{\mathrm{Born}}}{\mathrm{d}m}\cdot
\Delta m}\cdot \frac{1}{K}; \, \, 
K=\frac{\mathrm{d}\sigma_{\mathrm{Total}}^{\mathrm{vis}}/\mathrm{d}m}
{\mathrm{d}\sigma_{\mathrm{ISR}}^{\mathrm{vis}}/\mathrm{d}m} \, .
\label{ratio} 
\end{equation}
Detector acceptances and radiative corrections to the initial particles
are the same for
non-resonant and $ J/\psi $ contributions to
ISR  production of $ \mu^+ \mu^- \gamma $
and cancel in the ratio.
The total non-resonant cross section includes FSR contributions, which
we parameterize in terms of 
$K$, the ratio of the visible non-resonant total and ISR-only 
(FSR switched
off) cross sections. 
Using simulated events, we determine
$ K = 1.11\pm0.01$ (statistical error only) for our selection criteria.

The result of the fit is shown in Fig.~\ref{fit}.
We find $R = 18.94 \pm 0.44$ with
$ \chi^2 $ per degree of freedom $ \chi^2 / \nu=122/144$.
This fitted value  of $ R $ must be 
multiplied by 1.002 to correct for
non-resonant and resonant contributions from
$e^+e^-\to e^+e^-\gamma,\; \pi^+\pi^-\gamma,\; K^+K^-\gamma$.
The non-resonant cross section extracted from this measurement is
close to the value expected from simulation. 
Their ratio is
$0.968\pm0.016$. 
The quoted uncertainty includes a 0.4\% statistical error,
a 1\% statistical error from simulation,
and a 1.2\% uncertainty in luminosity.
We have not studied the systematic uncertainties on the efficiency
for the non-resonant process in detail, as 
most of these cancel
in $R$ and, hence, do not affect the measurement of the $J/\psi$ parameters.
The fitted value of $M_{J/\psi}$ is shifted from that
in the simulation by $-(1.6\pm0.3)$ MeV/$c^2$.
The fitted value of $\sigma_G  $ is $3.4\pm1.4$ MeV/$c^2$,
corresponding to an overall mass resolution 
$ ( \approx 14.5 \ \mbox{MeV}/c^2 ) $ 
3\% larger than that of the simulation.
The background slope $a$ 
corresponds to a 10\% change of the non-resonant
cross section in the mass range from 2.8 to 3.4 GeV/$c^2$.
The value of $b$ is consistent with zero, in agreement with
the Monte Carlo calculation.

As seen in Eq.~(\ref{ratio}), the ISR $J/\psi$ production
cross section is proportional to the product of $R$,
determined from fitting the data, and $ K $, determined
from Monte Carlo simulations. 
The primary sources
of systematic uncertainties for the product 
$ R \cdot K $ are summarized in Table~\ref{tab1}. 
Uncertainty in $K$ is caused by  different 
detection efficiencies for the pure ISR process
of $J/\psi$ production and the non-resonant
$e^+e^-\to \mu^+\mu^-\gamma$ process to which both ISR
and FSR amplitudes contribute.
We estimate the uncertainty due to data--Monte Carlo
differences by studying the stability of 
$R \cdot K$ for different selection criteria.
We vary the photon and muon angular selection criteria and the
muon momentum requirement over a wide range of values. 
While the value of  $K$ varies from 1.08 to
1.19, the maximum deviation from
our reference mean value $R \cdot K=21.03$ is only
1.3\%. 
Although this variation might be
a statistical fluctuation (at least in part),
we treat it as a systematic uncertainty associated with the
value of $ K $.
\begin{table}[t]
\begin{tabular}{lc}
\hline
statistical error of $K$ factor & 0.9\% \\
systematic error of $K$ factor & 1.3\% \\
background uncertainty & 0.5\% \\
simulation of $J/\psi$ line shape & 1.4\% \\
interference effect & 0.3\% \\
\hline
total                   & 2.2\% \\
\hline
\end{tabular}
\caption{The sources of systematic errors in $R \cdot K$.}
\label{tab1}
\end{table}

As described earlier, we use Monte Carlo simulations of
specific ISR and other processes to estimate the level
of non-resonant background to be less than 0.4\%.
We also use the data themselves to estimate 
this quantity.
We compare the fit results for data selected with the
standard selection criteria and  for data selected with additional
muon identification for one of the charged
particles. 
This reduces pion (kaon) contamination
by a factor of 9 (3).
From the difference in $R \cdot K$, we estimate
that the level of non-resonant background 
does not exceed 0.5\%.

The fit results do depend significantly on the model assumed for the
$J/\psi$ line shape.
The shape of the signal distribution varies with
the selection of the maximum allowed value of the $\chi^2$ from 
the one-constraint fit. Requiring lower values tends to reject events
with extra photons, thus reducing the fraction of events in the
low mass tail of the $ J/\psi $ peak.
The fraction of
$J/\psi$ events with mass less than $(M_{J/\psi}-0.1)$ GeV/$c^2$
changes from 2.4\% for no $\chi^2$ cut, to 0.4\% for
$\chi^2<5$.
Re-fitting data with different requirements on the value of $ \chi^2 $
does not change the result for $ R \cdot K $ significantly.
The maximum deviation of $ R \cdot K $ from our reference mean value, 
1.4\%, is taken as a systematic uncertainty.

We also consider an additional contribution
to the line-shape uncertainty by re-fitting
the data with
a model that does not include interference between
the non-resonant and $J/\psi$ production amplitudes.
The quality of this fit is good: $ \chi^2/\nu = 138/144 $.
As the data do not distinguish 
between the two models statistically, 
we take the 
difference in $R$, 0.3\%, as the corresponding systematic uncertainty.
The total systematic error for $K\cdot R$ is 
2.2\%, compared to the statistical error of 2.3\%.

The Born cross section for the process
$e^+e^-\to J/\psi\, \gamma \to \mu^+\mu^-\gamma$ can be evaluated
from Eq.~(\ref{ratio}). 
The non-resonant Born cross section in this formula
is calculated to be 
$\frac{\mathrm{d}\sigma_{\mathrm{ISR}}^{\mathrm{Born}}}{\mathrm{d}m}\cdot 
4 \mbox{ MeV}/c^2 = 101.0 \mbox{ fb}$.
Following the generally accepted practice~\cite{VP} of including
the vacuum polarization correction in the value of the electron width
$\Gamma_{ee}$, we multiply the pure Born cross section by
$1.042\pm0.002$. 
From 
$R\cdot K=21.03\pm0.49\pm0.47$ we calculate the cross section
$\sigma_{J/\psi}=2124\pm49\pm47 \mbox{ fb}$
and the product of the $J/\psi$ parameters
$$\Gamma_{ee}\cdot B_{\mu\mu}=0.3301\pm0.0077\pm0.0073 \mbox{ keV}.$$
From the PDG values~\cite{pdg} for $B_{ee}$ and $B_{\mu\mu}$,
which are dominated by those measured in 
$\psi(2S)\to J/\psi\pi^+\pi^-$ decays by the BES Collaboration~\cite{BES1},
we derive
the electronic and total widths of the $J/\psi$ meson,
$$\Gamma_{ee}=5.61\pm0.20\mbox{ keV},\:\:\:
\Gamma=94.7\pm4.4 \mbox{ keV} \, , $$
using the correlated errors reported by BES. 
The statistical and systematic 
uncertainties are combined in quadrature.
Our results agree with the previous world averages~\cite{pdg},
$\Gamma_{ee}=5.26\pm0.37\mbox{ keV}$
and $\Gamma=87\pm5 \mbox{ keV}$, but are more precise.
\vspace{1mm}\\
We are grateful for the excellent luminosity and machine conditions
provided by our \pep2\ colleagues, 
and for the substantial dedicated effort from
the computing organizations that support \babar.
The collaborating institutions wish to thank 
SLAC for its support and kind hospitality. 
This work is supported by
DOE
and NSF (USA),
NSERC (Canada),
IHEP (China),
CEA and
CNRS-IN2P3
(France),
BMBF and DFG
(Germany),
INFN (Italy),
FOM (The Netherlands),
NFR (Norway),
MIST (Russia), and
PPARC (United Kingdom). 
Individuals have received support from the 
A.~P.~Sloan Foundation, 
Research Corporation,
and Alexander von Humboldt Foundation.

\end{document}